\documentclass[authoryear,preprint,10pt,review]{elsarticle}
\usepackage{amssymb,amsmath,gensymb,multirow, threeparttable}
\usepackage{lineno}
\usepackage{booktabs}
\usepackage{hyperref}

\usepackage{graphicx}
\usepackage{wrapfig}
\usepackage{lscape}
\usepackage{rotating}
\usepackage{epstopdf}

\makeatletter
\def\bbordermatrix#1{\begingroup \m@th
  \@tempdima 4.75\p@
  \setbox\z@\vbox{%
    \def\cr{\crcr\noalign{\kern2\p@\global\let\cr\endline}}%
    \ialign{$##$\hfil\kern2\p@\kern\@tempdima&\thinspace\hfil$##$\hfil
      &&\quad\hfil$##$\hfil\crcr
      \omit\strut\hfil\crcr\noalign{\kern-\baselineskip}%
      #1\crcr\omit\strut\cr}}%
  \setbox\tw@\vbox{\unvcopy\z@\global\setbox\@ne\lastbox}%
  \setbox\tw@\hbox{\unhbox\@ne\unskip\global\setbox\@ne\lastbox}%
  \setbox\tw@\hbox{$\kern\wd\@ne\kern-\@tempdima\left[\kern-\wd\@ne
    \global\setbox\@ne\vbox{\box\@ne\kern2\p@}%
    \vcenter{\kern-\ht\@ne\unvbox\z@\kern-\baselineskip}\,\right]$}%
  \null\;\vbox{\kern\ht\@ne\box\tw@}\endgroup}
\makeatother

\journal{: arXiv} 
\begin{document}
\begin{frontmatter}
\title{Collective targeted migrations: a balancing act involving aggregation, group size and environmental clues: a simulation study}

\author[UFG]{Carlos Hernandez-Suarez}
 \ead{cmh1@cornell.edu}

\address[UFG]{Instituto de Innovaci\'on y Desarrollo, Universidad Francisco Gavidia, San Salvador, El Salvador}
  

\begin{abstract}
What is behind the \emph{wisdom of the crowds} described by  \cite{simons2004many}? It has been showed that insects may use gravitational fields to travel  \citep{DreyerDavid2018TEMF} and we may ask whether the use of gravitational fields is enough to secure the arrival of an individual to a relatively narrow spot thousands of kilometers away, as it is the case for example of monarch butterflies, which may travel 4,500 km to land in Mexico within an area of about 1/8 the size of Long Island. Here we show that if individuals budget a fraction of time to seek the target and the rest to maintain aggregation, the chances of landing within a narrow spot even under a weak signal are increased. Our model exhibits the existence of an equilibrium, the tradeoff that comes from maintaining group size, aggregation, and targeting. Whenever this balance is broken, the population may behave erratically.  If this strategic balance between targeting and aggregation is prevalent then the possibility of destabilizing may be used to control or regulate the spread of undesirable organisms that navigate under this balancing set of mechanisms. 

\end{abstract}

\end{frontmatter}


\section{Introduction}

Nature has plenty of examples where collective behavior is observed including populations of neurons  \citep{steur2009semi}, migrating cancer cells  \citep{10.1242/jcs.036525, C0IB00052C} and bacteria  \citep{Kato2018}, insects  \citep{deneubourg2002dynamics, anstey09}, human crowds  \citep{moussaid2009experimental} and whales  \citep{pomilla2005against}. The seminal work of the late Akira Okubo  \citep{okubo2013diffusion} opened the new frontier arising from possibility of exploring the role of mechanisms capable of capturing observed aggregation patterns of insects, birds or fish using physics-inspired modeling approaches  \citep{couzin05}.  A simple population model that increases the likelihood that individuals' tradeoffs involving targeting, aggregation and group size may play in reaching a desired target/location, is introduced.  

\section{Model description}
Our model has three parameters: $N$, the group size, $\theta$, the cohesion effort and $\sigma$, a measure of the innate ability of an individual to direct itself to a target. In this simplified model setting, it is assumed that each individual devotes a fraction of its handling time in analyzing external clues that impact the direction of travel with the remaining fraction committed to maintaining group cohesiveness (aggregation) for a fixed population size $N$. Our individual-based discrete-time model is used to explore the stochastic dynamics of individuals that live in a unit circle.  The population is initially placed at the center of this circle and it is assumed that it is targeting the point where the positive $x$-axis intersects the circle, that is, the point with coordinates $(1, 0)$. The information that each individual has on the target is modeled using a normal distribution, truncated in $(-\pi/2, \pi/2)$, with mean $0$ and variance $\sigma^2$. Individuals search for the target by traveling a distance $d$ in the direction $W$, a value taken from the ascribed distribution. Thus, $\sigma$ measures how well informed an individual is on the correct direction to travel, a `decision' that may change in response to clues emitted by the environment and the capability of each individual to process these signals, thus, $\sigma$ can be interpreted as the amount of noise present when calculating the direction of the next displacement. 

Model simulations suggests that there is an optimum group size $N$ and cohesion effort $\theta$ that get the work done and that this equilibrium is a function of the signal's weakness and the energy available to migratory species involving leaderless populations, `determined' to reach the target.

We use time-to-event simulations with a randomly chosen individual at each unit of time. It is then assumed that this individual either approaches another randomly chosen individual with probability $\theta$ or travels in a direction taken at random from the distribution $W$, with probability $1- \theta$. In both cases, the displacement is of size $d$. The process repeats until all individuals reach the circumference of the unit circle. Accuracy is defined as the distance of the mean position of the group to the target while precision is defined as the dispersion of individuals around the mean, both averages are computed at the end of the simulation. Successful migration requires high levels of accuracy and precision. Several movies capture simulations that show the displacement of the group for different parameter sets with jump size $d=1/100$ (see Supplementary Information). 

\section{Results}

For fixed $\sigma$,  it is assumed that the precision depends only on $\theta$ and not on $N$. Figure 1 shows the effect of varying $\theta$ on the precision and accuracy for $\sigma = 1$. In Figure 1.a we see precision increasing exponentially with $\theta$ while Figure 1.b shows accuracy decreases with $\theta$. The results in Figure 1.b may be the result of individuals reducing the time dedicated to seeking the target to a fraction $1- \theta$. Figure 1.b also shows that accuracy can be improved by increasing the group size for fixed $\sigma$ and $\theta$. 
The increase in accuracy as a function of group size and $\theta$ are illustrated in Figure 2, where we collect the mean average position of the group through 10 simulations for $\sigma = 3$. 

The random variable $Y$ ``Distance of an individual's final position to the target'' is used as a measure of efficacy in migration, whit $E[Y ] = 0$. Let $\bar{X}$ denote the random variable ``Average final position of the group'', thus, we can decompose $V [Y ]$ as $V [E[Y |\bar{X} ] + E[V [Y |\bar{X}]$. The first term in the decomposition of the variance is the accuracy, that is, variance of the expected position of an individual given $\bar{X}$, whereas the second term is the precision, the expectation of the variance of the position of an individual given $\bar{X}$. The first term can be estimated with the variance of the mean final position, $V [\bar{X}]$, whereas the second can be estimated by averaging the variances of the final positions, $E[S^2]$. $V [Y]$ is thus the sum of precision and accuracy and equals the mean square error (MSE). In general, an increase in group size increases accuracy and has no effect on the precision, while increasing $\theta$ has a mixed effect on MSE since it increases precision but reduces accuracy. 
Figure 3 shows the effect of $\theta$ on MSE for different group sizes for $\sigma=1$. We observe that for fixed $\sigma$ and fixed group size, increasing $\theta$ improves the MSE but only to the point where the effort dedicated to maintaining cohesion overcomes the effect of the group size, that is, the loss in accuracy that results from increasing $\theta$ does overcome the precision achieved that comes from increasing group size. 

If it is assumed that natural selection has shaped group size and group cohesion then reducing `optimal' group size, for instance, via an increase in hunting or predation or man-made causes, reduces the probability of arriving to the targeted destination for the group. We observe disorientation and erratic behavior. 
Figures 4.a and 4.b show two migration processes with $50$ simulations each, under parameters $N = 100, \sigma = 3$ and $\theta = 0.2$ In Figure 4.b the number of individuals is reduced to a $1/4$ of the original group size half the way to the target. Erratic behavior is more evident in this plot around the point of intervention, that is, when the population was reduced.

An expected fraction $\theta/2$ of displacements will result in a backward displacement of the mean of the group. Since an equivalent displacement in the opposite direction is required to recover, a fraction $\theta$ of the displacement energy is invested in maintaining cohesion. Simulations confirm this fact: if $t_\theta$ is the average number of steps needed to arrive to the target under a cohesive effort $\theta$, then simulations show that $(t_{\theta} - t_0)/t_\theta \approx \theta$ that is, the extra effort is spent in aggregation. Further, aggregation efforts increase the time to arrival in $t_0/(1 - \theta)$. It may be possible that in spite of the increase in time to reach the target, this `trick' may actually use energy more efficiently as long as the effort to remain aggregated consumes less energy than the energy required to process signals, environmental clues, leading to the target. 
Exploring the relationship between $\sigma$ and $\theta$ is important. Hernandez  \citep{Hernandez-SuarezCarlosM2016Aitk} extended the random walk in the line to include two individuals that spend a fraction of the time approaching each other, and showed that under certain conditions, aggregation increased the chances that both individuals arrived to the desired edge, and that the `doubling stakes' principle was behind the increase in success, since grouped individuals can be followed by the mean position that moves in jumps of $1/2$.

Here,  $\sigma$ and $\theta$ are assumed to be independent albeit it would be reasonable to assume that they are negatively correlated. That is, the larger the $\sigma$ the less the need to travel in group, one possibility may be $\theta(\sigma) = \sigma/(1 + \sigma)$. Clearly, different models of aggregation may yield different outcomes, for instance, traveling towards the center of mass. Our directional model is absolute, meaning the distribution $W$ is independent of the current position of an individual. There is a need to explore the effect of direction that depends on the current position. 

\section{Conclusion}

The wisdom of the crowds is based in the ``many wrongs'' principle  \citep{simons2004many} which states that the pooling of information from many inaccurate compasses yields a single more accurate compass. The results here suggest that increases in group size only account for increases in accuracy (average final position closer to the target) and not on increases in precision. Balancing the tradeoffs between $\theta$ and group size seem to be required to achieve any desired MSE. Here, we explore a navigation mechanism that boost precision and accuracy. This simplified model provides clues on the value of possible unexplored measures that can be used to avoid undesirable organized migrations such as those linked to cancer cells  \citep{10.1242/jcs.036525, C0IB00052C} and locusts  \citep{BAZAZI2008735} . Further, the possibility of increasing the ability of man-made devices to find the target, as nanorobots, that are conceived in principle as devices with a relatively low CPU capacity, may also benefit from the results and possibilities highlighted with this model.

\newpage
\section{Supplementary Information}

Movies with simulation of this process for several parameters are included at the Git repository:\\

\href{https://github.com/car-git/travelling-together}{https://github.com/car-git/travelling-together}\\

This is a description of the parameters used in each video. In all cases step size is $d=1/100$ and $\sigma=3$:


\begin{table}[h]
\begin{center}
\caption{List of files with simulations and parameters used.}
\begin{tabular}{@{}lcc@{}}
\toprule
\multicolumn{1}{c}{File} & N    & $\theta$ \\ \midrule
S1.mp4                   & 50   & 0        \\
S2.mp4                   & 50   & 0.2      \\
S3.mp4                   & 50   & 0.6      \\
S4.mp4                   & 200  & 0        \\
S5.mp4                   & 200  & 0.2      \\
S6.mp4                   & 200  & 0.6      \\
S7.mp4                   & 1000 & 0        \\
S8.mp4                   & 1000 & 0.2      \\
S9.mp4                   & 1000 & 0.6      \\ \bottomrule
\end{tabular}
\end{center}
\end{table}

\newpage
\section{Figures}

  \begin{sidewaysfigure}[htbp]
  \caption{Effect of the aggregation effort on precision (a) and accuracy (b). (a) shows the average dispersion of the final position of individuals: the variance of the final position of individuals was calculated and its average was calculated over 2000 simulations. The precision is independent of group size. (b) shows the variance of the mean final position: the mean of the final position of individuals was calculated and their variance was calculated over 2000 simulations, for different group sizes. This plot shows that for fixed $\theta$, the accuracy is increased with group size. Both plots use $\sigma=1$.}\label{Fig1}
\begin{center}
\includegraphics[width=22cm]{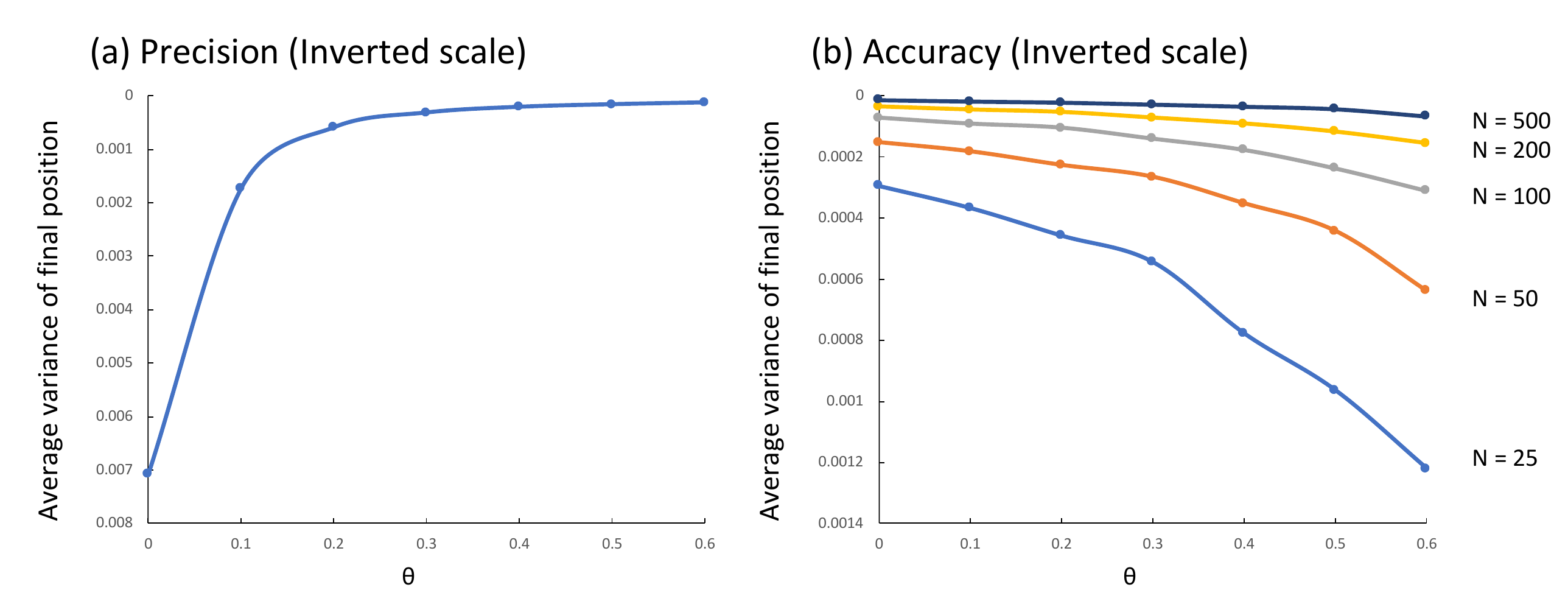}
\end{center}
\end{sidewaysfigure}

  \begin{sidewaysfigure}[htbp]
\caption{Effect of $N$ and $\theta$ on accuracy. Each line represents the path followed by the mean position of the group. For fixed $N$, the accuracy is reduced with $\theta$ and for fixed $\theta$, the accuracy is increased with $N$.}\label{Fig2}
\begin{center}
\includegraphics[width=20cm]{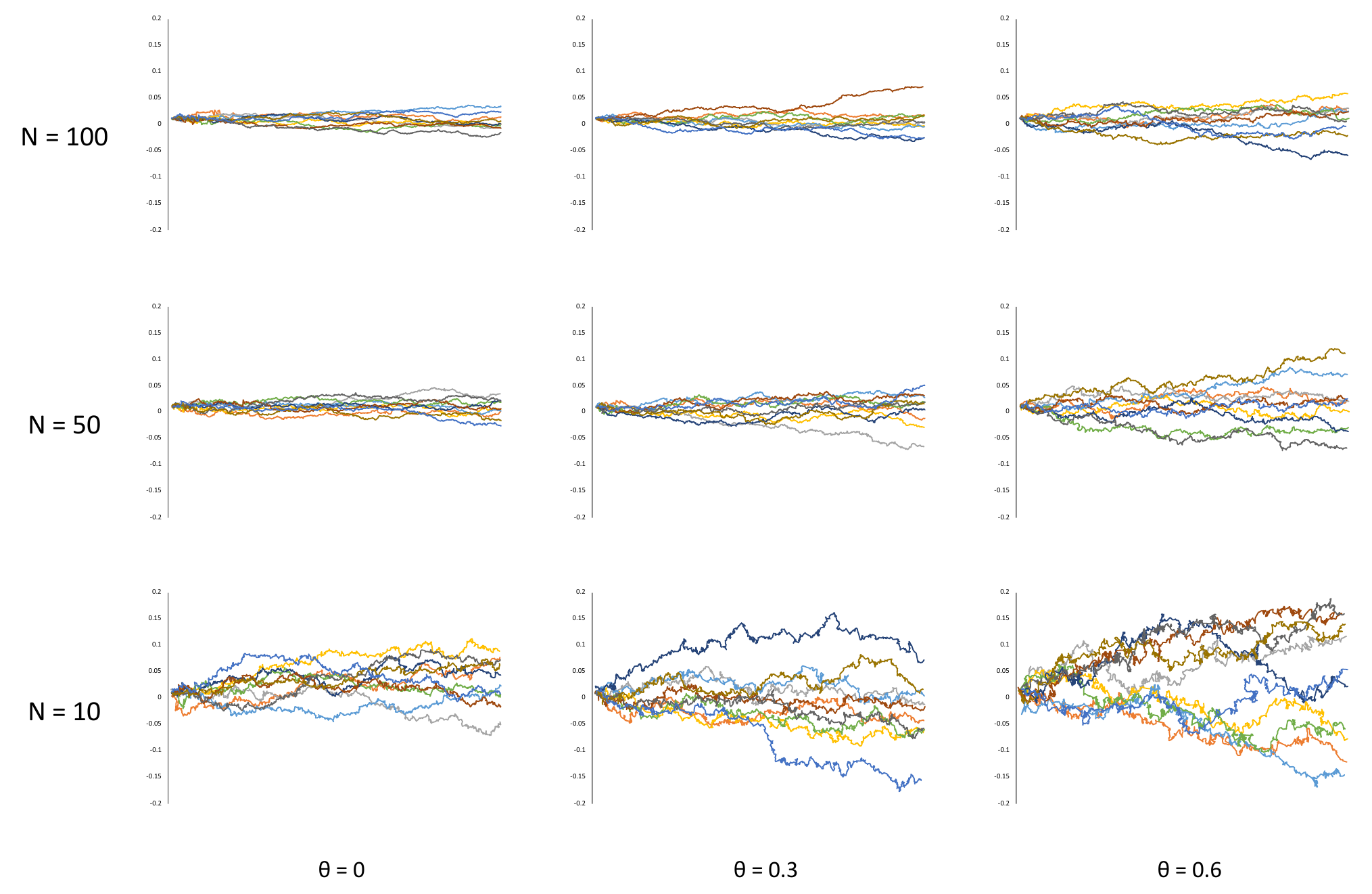}
\end{center}
\end{sidewaysfigure}

  \begin{figure}[htbp]
\caption{Mean square error, MSE = accuracy + precision as a function of $\theta$ for different group sizes. For this plot $\sigma=1$.  In the range studied, at least for $N \leq100$ the MSE reduces up to a point where it starts increasing again. The larger $\theta$ the less time an individual invests in locating the target.}\label{Fig3}
\begin{center}
\includegraphics[width=14cm]{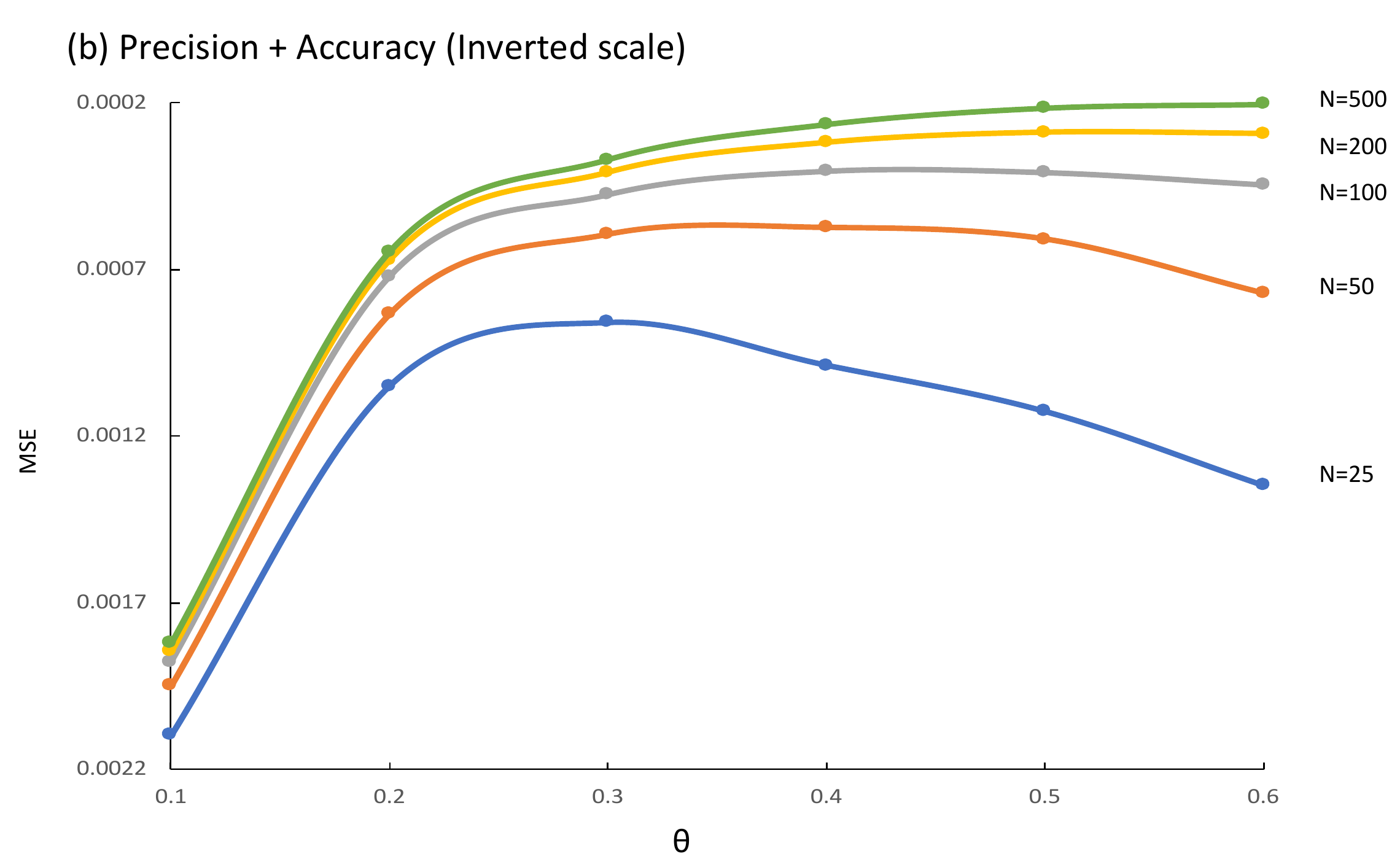}
\end{center}
\end{figure}

  \begin{sidewaysfigure}[htbp]
\caption{Effect of a disruption in the migration process. Plot (a) shows the paths of the mean position of the group across 50 simulations. The parameters used are $N=100$, $\sigma=3$ and $\theta=0.2$. The simulations in (b) show an identical process but with a disruption consisting in the sudden reduction of the group to a quarter of its original size half the way to the target. The variance of the mean position of the group increases since individuals ignore the group has been reduced and keep moving according a $\theta$ that is optimal for a larger group size.}\label{Fig4}
\begin{center}
\includegraphics[width=20cm]{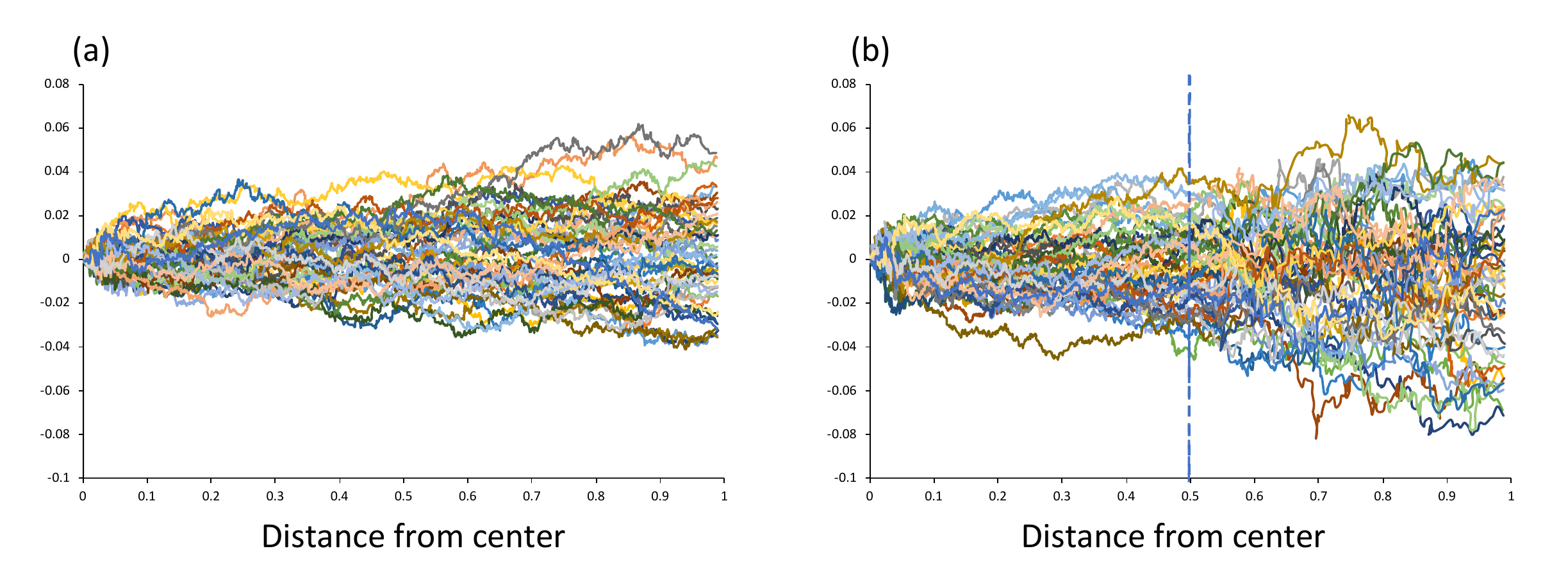}
\end{center}
\end{sidewaysfigure}

\end{document}